\input{aipcheck}

\documentclass[
    ,final            
  ]
  {aipproc}

\layoutstyle{6x9}
\newcommand{\be}{\begin{equation}}
\newcommand{\ee}{\end{equation}}
\newcommand{\bea}{\begin{eqnarray}}
\newcommand{\eea}{\end{eqnarray}}

\newcommand{\rr}[4]{#1, {\it #2 \/}{\bf #3} #4}

\begin{document}

\title{NLO BFKL tests using the proton structure function $F_2$ measured at HERA}


\author{C. Royon}{
  address={Service de physique des particules, CEA/Saclay,
  91191 Gif-sur-Yvette cedex, France \\ Fermilab, Batavia, USA}}



\begin{abstract}
We propose a phenomenological study of the next-to-leading 
Balitsky-Fadin-Kuraev-Lipatov
(BFKL) approach applied to the data on  the proton structure 
function $F_2$ 
measured at HERA in the small-$x_{Bj}$ region. 
\end{abstract}

\maketitle


\section{Introduction}

Precise phenomenological tests of QCD evolution equations are one of the 
main 
goals of deep inelastic scattering phenomenology. For  the 
Dokshitzer-Gribov-Lipatov-Altarelli-Parisi
(DGLAP) evolution in  $Q^2$
\cite{dglap}, it has 
been possible 
to test it in various ways with  NLO (next-to-leading $\log Q^2$) and 
now NNLO accuracy and 
it works quite well in a large range of $Q^2$ and $x_{Bj}.$
Testing precisely the Balitsky-Fadin-Kuraev-Lipatov (BFKL) 
evolution in energy \cite{bfkl} (or  $x_{Bj}$) beyond leading order appears 
more difficult.

The first experimental  results from 
HERA  confirmed the existence of a strong rise of the proton structure 
function $F_2$ 
with energy which, in the BFKL framework, can be   well described by a simple 
(3 parameters) LO-BFKL fit \cite{old}. 
The main issue of Ref.\cite{old} was that not only the rise with energy but 
also the scaling violations  observed at small ${x_{Bj}}$ are encoded
in the BFKL framework through the $Q^2$ variation of the effective 
anomalous dimension. However one was led  \cite{old} to introduce an 
effective but unphysical  value of the  strong coupling constant  
$\alpha \sim .07-.09$ 
instead of 
$\alpha \sim .2$ in the $Q^2$-range considered  for  HERA small-${x_{Bj}}$ 
physics, revealing the need for NLO corrections. Indeed, the 
running of the strong coupling constant is not taken into account. 

In fact,  the  theoretical task of computing these   corrections 
appears to be quite hard. It is now in good progress but  
still under completion. For the BFKL kernel, they have been 
calculated after much efforts \cite{next}. 
It was realized \cite{salam} 
that the main problem comes from the existence of spurious singularities 
brought together with the NLO corrections, which ought to be cancelled 
by an appropriate resummation at all orders of the perturbative 
expansion \cite{next, autres, lipatov}, resummation
required by consistency with  the QCD renormalization group. 

\section{NLO BFKL phenomenology}

\subsection{Saddle point approximation}
Following the successful BFKL-LO parametrisation of the proton structure
$F_2$ at HERA, we perform the same saddle point approximation as at LO using
$\chi_{NLO}$ given by resummed NLO BFKL kernels \cite{us}.
\begin{eqnarray}
F_2 = C e^{\alpha_{RGE} \chi_{eff} (\gamma_{c},
\alpha_{RGE}) Y} \left( Q^2/Q_0^2 \right)^{\gamma_{c}} 
e^{- \frac{log^2 (Q/Q_0)}{2 \alpha_{RGE} \chi''_{eff}
(\gamma_{c},\alpha_{RGE}) Y}}
\end{eqnarray} 
where $\gamma_C$ and $\chi_{eff}$ come directly from the properties of the NLO
BFKL equation if the small-x structure function is dominated by the perturbative
Green function:

\begin{eqnarray}
\frac{d \chi_{eff}}{ d \gamma }(\gamma_C, \alpha_{RGE} (Q^2)) = 0 ~~~;~~~~
\chi_{eff} (\gamma, \alpha_{RGE} ) = \omega (\gamma, \alpha_{RGE}) /
\alpha_{RGE}.
\end{eqnarray} 

Instead of getting a 3-parameter formula like at LO (normalisation, $\alpha_S$,
and $Q_0$), we get only two free parameters at NLO since the value of $\alpha_S$
and its $Q^2$ evolution are imposed by the renormalisation group equations
(RGE). The delicate aspect of the problem comes for the fact that $\chi$ is now 
scheme dependent. 

\subsection{Strategy for NLO fits}
The strategy for BFKL-NLO is the following \cite{us}:
\begin{itemize}
\item The first step is the knowledge of $\chi_{NLO}(\gamma, \omega, \alpha)$
from the BFKL equation and different resummation schemes
\item The second step is to use the implicit equation $\chi (\gamma, \omega) = \omega / \alpha$
to compute numerically $\omega$ as a function of $\gamma$ for different
schemes and values of $\alpha$
\item The third step is to determinate numerically the saddle point values $\gamma_C$
as a function of $\alpha$ as well as the values of $\chi$ and 
$\chi ''$
\item The fourth step is to perform the BFKL-NLO fit to HERA $F_2$ data with two
free parameters $C$ and $Q_0^2$
\end{itemize}
Details about the numerical results can be found in Ref. \cite{us}. 

The results of the NLO BFKL fit
to the H1 and ZEUS data \cite{data} using the saddle point approximation for two
different schemes (CSS and S3, see Ref \cite{salam, autres}) are given in Fig 1 
where the 
data over theory ratio is displayed.


\begin{figure}[htb]
\includegraphics[width=12.5cm,clip=true]{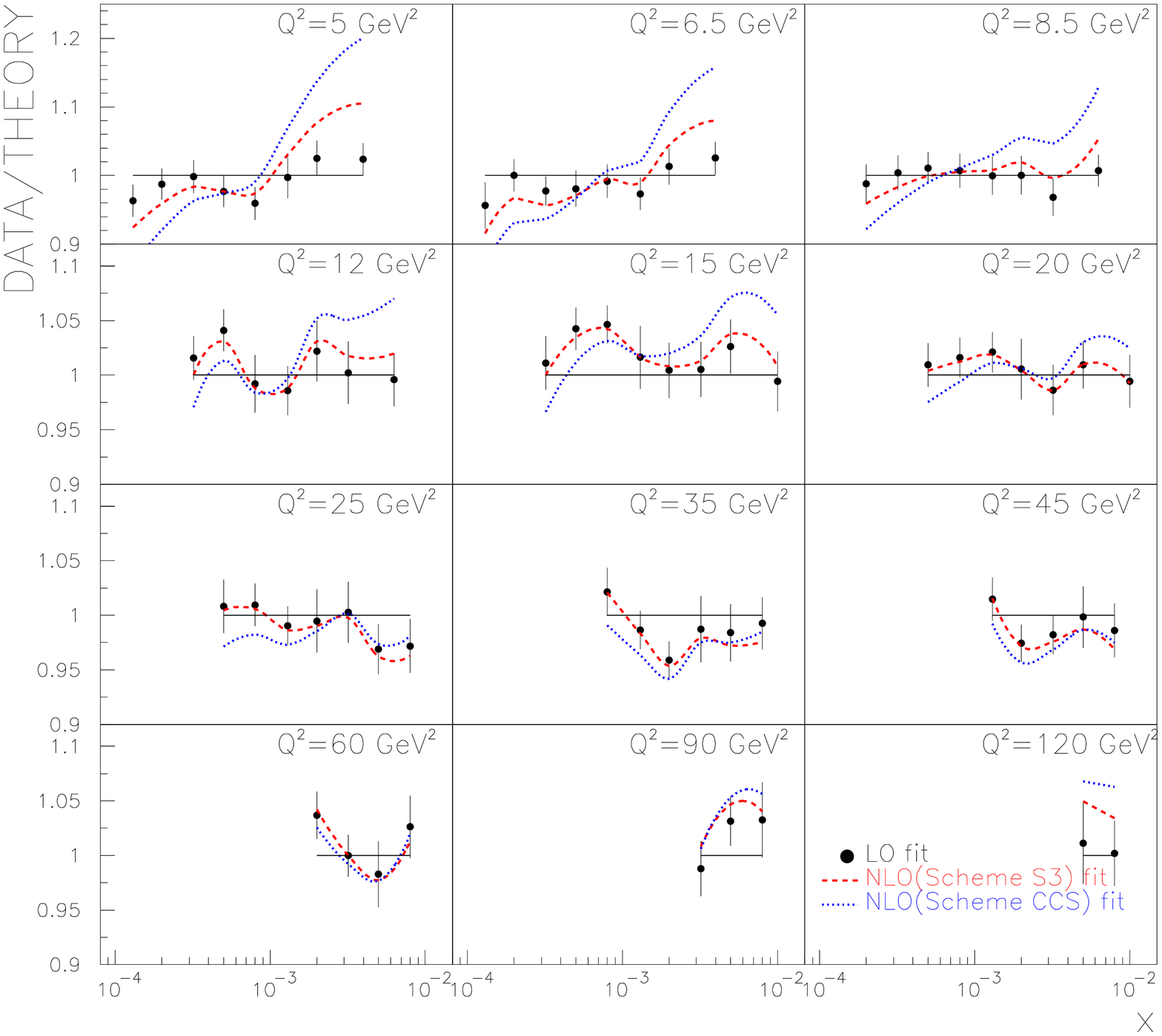}
\caption{Data/Theory ratios for the proton structure function $F_2$.
  The points show the results of the LO fit as a reference. The dashed (resp. 
  dotted) lines show the results of the BFKL-NLO fits including the S3 
  (resp. CCS) resummation
  scheme.}
\end{figure}

We see a big dicrepancy between data and theory especially at lower $Q^2$.
To understand further the reason of that discrepancy, we performed an analysis
in the Mellin space where the formulation of the BFKL NLO resummed kernels is
easier.

\section{Analysis in Mellin space}
In this section we want to analyze in more detail the features of the 
BFKL 
parametrizations and in particular the reasons of the still quantitatively 
unsatisfactory 
results of the NLO fits. For this sake, it is important to come back to 
the 
key ingredient of our analysis, i.e. the dominance of the hard Pomeron 
singularity expressed by the relation (2). 
Equality  (2)  can be checked at NLO using the GRV98 
\cite{GRV98}, MRS2001 \cite{MRS2001}, CTEQ6.1 \cite{CTEQ6.1} and ALLM 
\cite{ALLM}
parametrisations. These four parametrisations give a fair description of 
the 
proton
structure functions measured by the H1 and ZEUS collaborations over a 
wide 
range of $x_{Bj}$ and $Q^2$, as well as fixed target experiment data. The 
three first
parametrisations correspond to a DGLAP NLO evolution whereas the ALLM 
one
corresponds to a Regge analysis of proton structure function 
data.

We notice in Fig.2 that the linear property of relation 
(2), namely for $\chi_{eff} (\gamma^*,\alpha_{RG})$ as a 
function of $\omega$ 
is well verified. We indeed  can describe
the GRV and MRS parametrisations using a linear fit with a good 
precision. However 
the predicted zero at the origin   $\omega =0$ is not obtained, even if the 
value at 
the origin 
remains small. The fit does not go through the origin and we would need to 
add a constant term to the linear fit formula, and the slope is not equal
to $\alpha$. Small but sizeable 
effects give phenomenological deviations from the expected theoretical 
properties of the NLO kernels.

\begin{figure}[htb]
\includegraphics[width=12.5cm,clip=true]{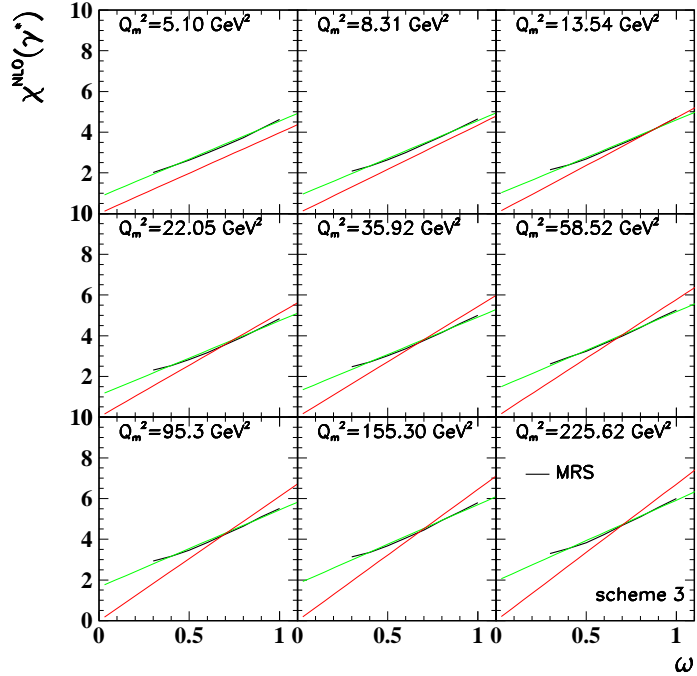}
\caption{Test of $\chi(\omega,Q^2) $ for scheme $S3$. The 
result for the  
MRS parametrization is shown in black in the different bins in $Q^2$  together 
with a linear fit, and the expectation if formula (2) is fulfilled. We notice
the discrepancy between the MRS result and formula (2). }
\end{figure}

\section*{Acknowledgments}
There results come from a fruitful collaboration with R. Peschanski and L.
Schoeffel.



 



\end{document}